\documentstyle[12pt,axodraw]{article}

\newcommand{\eq}[1]{Eq.(\ref{#1})}

\newcommand{\dirac}[1]{ {{#1}\!\!\!{\slash}} }
\newcommand{\bm}[1]{\mbox{\boldmath$#1$}}

\begin{document}

\begin{center}
{\bf NUCLEAR SHADOWING IN NEUTRINO DEEP INELASTIC SCATTERING
}\footnote{
      Based on the talk given at XIV International Seminar
{\em `Relativistic Nuclear Physics And Quantum Chromodynamics' }
       Dubna, August 17-22, 1998. To be published in proceedings.
}

\vskip 5mm
S. A. Kulagin

\vskip 5mm

{\small
{\it
        Institute for Nuclear Research,
        Russian Academy of Sciences,
        Moscow, Russia
}
\\
{\it
E-mail: kulagin@ms2.inr.ac.ru
}}
\end{center}

\vskip 5mm

\begin{center}
\begin{minipage}{150mm}
\centerline{\small\bf Abstract}
{\small
We discuss effect of nuclear shadowing in neutrino deep-inelastic
scattering in terms of non perturbative parton model.
We found that for small Bjorken $x$ and large $Q^2$
the structure function $F_3$ is shadowed in nuclei
about two times as stronger as $F_2$.
The underlying reason and phenomenological aspects of this
observation are discussed.
}
\end{minipage}
\end{center}



\section{Introduction}

The small $x$ physics is an important topic of high energy studies nowadays.
We note in this respect that data is usually collected for nuclear targets
and it is important to consider nuclear effects in deep inelastic scattering
(DIS).
The systematic reduction of the nuclear structure functions $F_2^A/A$
with respect to the nucleon $F_2^N$ has been established
experimentally at CERN \cite{nmc} and Fermilab \cite{e665}
in the region of small Bjorken $x$,
so called nuclear {\em shadowing} phenomenon.
This phenomenon has been extensively discussed in the literature
within different theoretical approaches
(for a review see ,e.g., ref.\cite{survey}).


Apart from $F_2$, it is important of course to study the nuclear effects
in the other observed structure functions.
An interesting object in this respect is the structure function $F_3$
which is measured in (anti)neutrino DIS.
We recall that at large momentum transfer $F_3$
describes the valence quark distribution in the target.
The main concern of the present paper is to study the nuclear effects
in neutrino DIS in the small $x$ region and compare the shadowing effect
for the structure functions $F_2$ and $F_3$,
i.e. for the sea and the valence quark distributions.

For the discussion of nuclear physics at DIS it is convenient
to choose the target rest frame.
To clarify the space-time picture
of DIS in this reference frame
it is convenient to argue using the time-ordered (`old')
perturbation theory.
We recall that
DIS cross section is determined by the imaginary part of the Compton
scattering amplitude of the virtual intermediate boson from the target.
Two typical time-ordered diagrams which
contribute to the Compton nucleon scattering amplitude are drawn in
Fig.1.

%
\begin{center}
\begin{picture}(360,50)
%
\SetWidth{0.5}
\ArrowLine(20,10)(45,30)
\ArrowLine(45,30)(75,30)
\ArrowLine(75,30)(100,10)
\DashLine(32,0)(32,40){3}\Text(32,50)[]{$t_1$}
\DashLine(60,0)(60,40){3}\Text(60,50)[]{$t_2$}
\Photon(20,40)(45,30){2}{5}
\Photon(75,30)(100,40){-2}{5}
\Vertex(45,30){2}
\Vertex(75,30){2}
\Line(20,10)(100,10)
\Line(20,5)(100,5)
\Text(60,-10)[]{$a$}
\SetWidth{1.5}
\ArrowLine(0,0)(20,5)
\LongArrow(100,5)(120,0)
\SetWidth{0.5}
\GOval(20,10)(10,7)(0){0.9}
\GOval(100,10)(10,7)(0){0.9}
%
%
\SetWidth{0.5}
\ArrowLine(260,10)(220,35)
\ArrowLine(220,35)(340,35)
\ArrowLine(340,35)(300,10)
\DashLine(235,0)(235,40){3}\Text(235,50)[]{$t_1$}
\DashLine(280,0)(280,40){3}\Text(280,50)[]{$t_2$}
\Photon(200,40)(220,35){2}{4.5}
\Photon(340,35)(360,40){-2}{4.5}
\Vertex(220,35){2}
\Vertex(340,35){2}
\Line(260,15)(300,15)
\Line(260,12.5)(300,12.5)
\Line(260,10)(300,10)
\Line(260,7.5)(300,7.5)
\Line(260,5)(300,5)
\Text(280,-10)[]{$b$}
\SetWidth{1.5}
\ArrowLine(240,0)(260,5)
\LongArrow(300,5)(320,0)
\SetWidth{0.5}
\GOval(260,10)(10,7)(0){0.9}
\GOval(300,10)(10,7)(0){0.9}
\end{picture}
\\{\small FIG.1}\\
\vspace{10pt}{\small
Time-ordered diagrams $(t_1<t_2)$ for the Compton scattering.}
\end{center}

\noindent
Fig.1a describes the scattering process which goes via absorption
and re-emission of the virtual boson by a quark bound in the target.
Fig.1b shows the process where
the virtual boson first decays into a $q\bar q$-pair which then
interacts with the target.
By comparing the energy denominators of these two diagrams one can
observe that at small $x$ the diagram in Fig.1b dominates over the
one in Fig.1a.  This observation leads to the qualitative space-time
picture of the Compton scattering at small $x$
which consists from two time-ordered processes.  First,
the virtual boson splits into a $q\bar q$ (or hadronic) fluctuations
before arrival at the target.  Then these states propagate and
interact with the target.
A typical propagation length
of a $q\bar q$ fluctuation can be estimated from the uncertainty principle
and at high $Q^2$ is of order $L\sim (2Mx)^{-1}$ with $M$ the nucleon mass%
\footnote{
$L$ is also a typical time between absorption and emission of the
virtual boson in Fig.1}
\cite{IoKhLi84}.
For $x\ll 0.1$ the $q\bar q$ fluctuation lives long enough to develop a chain
of multiple interactions with bound nucleons. This will lead
to nuclear shadowing effect which we discuss in detail in Sec.\ref{shadow}.
Before we come to the discussion of nuclear modifications in the DIS,
in Sec.\ref{framework}
we set up the model for the DIS from the isolated nucleon
and fix some parameters of the model by comparing to data.

\section{DIS in non perturbative parton model}
\label{framework}

We consider the neutrino and antineutrino charged-current DIS from the nucleon.
It is supposed that $Q^2$ is sufficiently high to neglect the target
mass effects. We neglect also possible final state interaction effects
and focus on the leading
twist contribution to the DIS which is given by the handbag diagram of Fig.2
(note that this diagram is treated in terms of a covariant approach which
includes of course all time-ordered processes shown in Fig.1).

%
\begin{center}
\begin{picture}(300,100)
\SetWidth{0.5}
\ArrowLine(15,25)(30,65)
\ArrowLine(30,65)(80,65)
\Text(30,50)[]{$k$}
\Text(55,55)[]{$k+q$}
\ArrowLine(80,65)(95,25)
\SetWidth{1.5}
        \ArrowLine(0,5)(15,25)
        \Text(10,0)[]{$P$}
        \LongArrow(95,25)(110,5)
\SetWidth{0.5}
\Photon(5,95)(30,65){2}{7}
        \Vertex(30,65){2}
        \LongArrow(15,95)(22.3,85)
\Photon(80,65)(105,95){-2}{7}
        \Vertex(80,65){2}
        \LongArrow(87,85)(95,95)
\GOval(55,25)(15,50)(0){0.9}
\Text(55,25)[]{$\chi(k,P)$}
\Text(150,25)[]{$+$}
\ArrowLine(205,25)(220,65)
\ArrowLine(220,65)(270,65)
\Text(220,50)[]{$k$}
\Text(245,55)[]{$k-q$}
\ArrowLine(270,65)(285,25)
\SetWidth{1.5}
        \ArrowLine(190,5)(205,25)
        \Text(200,0)[]{$P$}
        \LongArrow(285,25)(300,5)
\SetWidth{0.5}
\Photon(220,95)(270,65){2}{10}
        \Vertex(270,65){2}
        \LongArrow(230,96)(240,90)
\Photon(220,65)(270,95){-2}{10}
        \Vertex(220,65){2}
        \LongArrow(250,90)(260,96)
\GOval(245,25)(15,50)(0){0.9}
\Text(245,25)[]{$\chi(k,P)$}
\end{picture}
\\{\small FIG.2}
\\
\vspace{10pt}{\small
Compton scattering amplitude from the nucleon to the leading order in $Q^2$.}
\end{center}
In the axial gauge the Compton amplitude for the scattering of virtual
$W$-boson from the target reads
(to be specific we consider the Compton scattering of $W^+$-boson):
\begin{eqnarray}
\label{IA}
T_{\mu\nu}(P,q) &=& -i\int \frac{d^4k}{(2\pi)^4}
               {\rm Tr}\left[
\chi_d(k,P) \gamma_{\mu}(1+\gamma_5)
(\dirac{k}+ \dirac{q}+i\epsilon)^{-1}
\gamma_{\nu}(1+\gamma_5) + {}\right. \nonumber \\
 && \left.
        \chi_u(k,P)\gamma_{\nu}(1+\gamma_5)
        (\dirac{k} - \dirac{q}+i\epsilon)^{-1}
         \gamma_{\mu}(1+\gamma_5)\right] + \ldots,
\end{eqnarray}
where $q$ is momentum transfer and $\chi_a(k,P)$ is
the propagator of the quark fields of kind $a$ in the target
with momentum $P$,
\begin{eqnarray}
\label{chi}
    \chi_a(k,P)= -i\int d^{4}\xi\: e^{ik\cdot\xi} \langle
    P|T\left(\psi_a(\xi)\bar{\psi}_a(0)\right)|P \rangle  .
\end{eqnarray}
The dots in \eq{IA} denote contributions from the $s$- and $c$-quarks which
repeat the first two terms with the substitutions $u\to c$ and $d\to s$.

The structure functions are given by the imaginary part of the
Compton amplitude and can be found from Eq.(\ref{IA}) by applying
appropriate projection operators.
In the leading order in $1/Q^2$ we find that the structure functions
$F_2$ and $F_3$ are given in terms of the quark light-cone
distributions by the standard parton model equations
(neglecting a small effect due to Cabibo mixing angle),
\begin{eqnarray}
\label{F2IA}
F_2^\nu(x) &=& 2x\left(d(x)+\bar u(x) +s(x) +\bar c(x)\right) ,\\
F_3^\nu(x) &=& 2\left(d(x)-\bar u(x) +s(x) -\bar c(x)\right).
\end{eqnarray}
where the parton distributions are expressed in terms of the quark
propagator as follows,
\begin{eqnarray}
\label{f}
f_a(x) &=& -i\int \frac{d^4k}{(2\pi)^4}
\frac{{\rm Tr}\left(\dirac{q}\chi_a(k,P)\right)}{2P\!\cdot\! q}\,
                \delta\left(x-\frac{k\!\cdot\! q}{P\!\cdot\! q}\right) ,\\
q_a(x) &=& f_a(x),\  \bar q_a(x) = -f_a(-x) .
\label{qqbar}
\end{eqnarray}
Note that the quark and antiquark distributions come from the  direct
and the crossed terms of the Compton amplitude respectively.
It follows from Eq.(\ref{IA}) that
the structure function $F_1$ is not independent; it
is given by the Callan-Gross relation, $F_2(x)=2x\,F_1(x)$.

In QCD scaling is violated and the parton distributions depend also
on $Q^2$. The $Q^2$ dependence comes through the dependence of the
quark correlator on the normalization point $\mu^2$ which is taken
to be $Q^2$ in our discussion.

For our purposes it will be convenient to write the parton distributions
as the integrals over the invariant mass spectrum of
the spectator states \cite{LPS}.
To this end we assume the analytic properties
of the quark correlator (\ref{chi})
to be similar to the ones of hadronic amplitudes,
i.e.
it is an analytic function of the invariant variables
 $s=(P-k)^2$, $u=(P+k)^2$ and
$k^2$.  For real $s$ and $u$ the quark amplitude has a right-hand  cut in
the variable $s$, a left-hand cut in the variable $u$ and singularities at
$k^2>0$.  In order to make use of these analytical properties of
in the loop integral (\ref{f}), it is convenient to parameterize the loop
momentum $k$ in terms of the light-cone or
the Sudakov variables.
Acting along these lines one finds
that the distribution function $f(x)$ vanishes outside the physical interval
$|x|\le 1$. For $0\le x\le 1$ the distribution function $f(x)$
(or the quark distribution $q(x)$)
is given by a dispersion integral along the cut in the
$s$-channel, while for $-1\le x\le 0$ the $u$-channel cut is relevant.
We separate from the quark correlator (\ref{chi}) external quark
propagators and write the result in terms of
the quark-nucleon, $T_{qN}$, and the antiquark-nucleon, $T_{\bar qN}$,
scattering amplitudes.  In terms of these amplitudes the small $x$ part
of the quark and
antiquark parton distributions read as follows
(to simplify notations we drop here explicit dependence on the
quark kind $a$)
\begin{eqnarray}
\label{qII}
q(x) &=& \frac{N_c}{(2\pi)^3}
    \int ds\int^{k^2_{{\rm max}}(s,x)}dk^2
        \:\frac{{\rm Im} T_{\bar qN}(s,k^2)}{(k^2-m_q^2)^2}
    \left(-x+\frac{m_q^2-k^2}{s-M^2-k^2}\right) ,\\
\label{qbarII}
\bar q(x) &=& \frac{N_c}{(2\pi)^3}
    \int du\int^{k^2_{{\rm max}}(u,x)}dk^2
        \:\frac{{\rm Im} T_{qN}(u,k^2)}{(k^2-m_q^2)^2}
    \left(-x+\frac{m_q^2-k^2}{u-M^2-k^2}\right) .
\end{eqnarray}
Here $N_c$ is the number of colors, $m_q$ is the quark mass.
The integrations run over the spectrum of intermediate states
and over the quark four-momentum squared $k^2$
with $k^2_{\rm max}(s,x)=x(M^2-s/(1-x))$ being the kinematical maximum
of $k^2$ for the given $s$ and $x$.
We note that the quark distribution $q(x)$ is determined by the
$\bar qN$-scattering amplitude, in accordance with intuitive discussion
based on time-ordered diagrams of Fig.1.
For the antiquark distribution $\bar q(x)$ the $qN$-amplitude
is relevant.
The reader can find the derivation
of Eqs.(\ref{qII},\ref{qbarII}) in ref.\cite{KPW94}.

Eqs.(\ref{qII},\ref{qbarII}) establish the correspondence between
the $x$-dependence of parton distributions
and the region of invariant masses of the residual quark system.
The basic assumption of the non-perturbative parton model is that
the parton amplitudes vanish at large virtualities so that
the integrals in Eqs.(\ref{qII},\ref{qbarII})
are finite and saturated in the region
of finite $k^2$. At small $x$ the $k^2$ is finite even for
large $s\sim M^2/x$ and therefore the parton distributions
are sensitive to the large-$s$ part of the spectral densities.

In ref.\cite{KPW94} (see also \cite{KMWW96}) we have considered
a simple model for the spectrum of the residual quark system
which incorporates two scattering mechanisms
shown in Fig.1. It was assumed that
the spectrum of spectator states is divided into low-mass
and high-mass parts sharply at some separation parameter $s_0$. The
physical origin of the
low-mass part of the spectrum is due to the Compton scattering of the
intermediate boson from bound quarks in the target, Fig.1a.
This process is of major importance at intermediate and
large $x$.
The high-mass part of the spectrum ($s>s_0$) gives rise to
the low-$x$ part of the structure functions.
In this region the spectrum is
dominated by the process shown in Fig.1b.

In order to calculate the quark and anti-quark distributions one has
to specify the amplitudes $T_{qN}$ and $T_{\bar qN}$. We follow here
ref.\cite{KPW94} and extract the quark-nucleon amplitudes from
the observed high-energy hadron scattering amplitudes. The idea is to
approximate the amplitudes $T_{qN}$ and $T_{\bar qN}$ by the
constituent quark--target scattering amplitudes which we obtain from
the additive quark model of high-energy scattering.
We are concerned here with the scattering from the isoscalar target.
The isoscalar parts of the (anti)proton-nucleon forward scattering
amplitudes,
$T_{pN}=(T_{pp}+T_{pn})/2$ and $T_{\bar pN}=(T_{\bar pp}+T_{\bar pn})/2$,
can be written in terms of the (anti)quark-nucleon
amplitudes as,
\footnote{
In fact if the isospin is the exact symmetry one finds
$T_{uN}=T_{dN}=T_{qN}$ and $T_{\bar uN}=T_{\bar dN}=T_{\bar qN}$.
We have used this observation implicitly
by writing Eq.(\ref{Tav}) in terms of
only two isoscalar quark-nucleon amplitudes.
}
\begin{eqnarray}
T_{pN}(S)/S &=& \left\langle
T_{qN}(y S,k^2)/y S
\right\rangle,  \nonumber\\
T_{\bar pN}(S)/S &=& \left\langle
T_{\bar qN}(y S,k^2)/y S
\right\rangle.
\label{Tav}
\end{eqnarray}
Here $S$ is the nucleon-nucleon invariant mass squared, $k$ is the
four-momentum of a constituent quark in the target and $y$ is a
fraction of the target light-cone momentum carried by the constituent
quark.  The averaging is taken over the quark distributions in the
proton which is normalized to the number of quarks.

An approximate solution of Eqs.(\ref{Tav}) can be obtained
by evaluating
the $qN$ and $\bar qN$ amplitudes at averaged values of $y$ and $k^2$
for constituent quarks in the proton,
$\bar y\sim 0.2$, $\overline{k^2}\sim -0.3\,$GeV$^2$.
The $s$-dependence of the (anti)quark amplitudes
is then fixed by the energy dependence of the observed
$pN$ and $\bar pN$ scattering amplitudes.
In the Regge approach these amplitudes can be approximated by the exchange
of the Pomeron $P$ and two Regge poles corresponding to the scalar $(f)$
and the vector meson $(\omega)$ trajectories \cite{Co},
$T_{pN}=P+f-\omega$, $T_{\bar pN}=P+f+\omega$.
Eqs.(\ref{Tav}) thus determine the quark--nucleon Regge poles
residues in terms of the nucleon--nucleon ones.
The exchange of the Regge trajectory with the intercept $\alpha_R$
leads to the imaginary parts of the amplitudes rising as $s^{\alpha_R}$.
The Pomeron exchange dominates
both $q(x)$ and $\bar q(x)$ at small $x$
which rise asymptotically as $x^{-\alpha_P}$, as one can
see from Eqs.(\ref{qII},\ref{qbarII}).
The Pomeron and the scalar $f$-Regge pole cancel out
in the difference $q-\bar q$ (the valence quark distribution).
Note however that the valence quark distribution remains finite at small
$x$ with the asymptotics determined by the $\omega$ Regge trajectory,
$q(x)-\bar q(x)\sim x^{-\alpha_\omega}$.
In the numerical estimates we use the parameters of
best fit of ref.\cite{DL92}
with the Regge poles intercepts $\alpha_P=1.08$
and $\alpha_f=\alpha_\omega=0.55$.

In the model of ref.\cite{KPW94} it is assumed that the coupling of
the Regge poles to quarks is independent of its kind,
and thus the different parton distributions are described
by two amplitudes $T_{qN}$ and $T_{\bar qN}$.%
\footnote{
We note in this respect that the parton distributions depend on their masses,
as one can see from Eqs.(\ref{qII},\ref{qbarII}).
Therefore the contribution from heavy quarks is suppressed
by their large masses.}
In order to have the integrals in
Eqs.(\ref{qII},\ref{qbarII}) convergent the amplitudes should fall
sufficiently fast for large quark virtualities $k^2$.
We assume the factorized $s$- and
$k^2$-dependencies for every Regge pole in the quark-nucleon
amplitude which we parameterize in a simple form
$(1-k^2/\Lambda_R^2)^{-n_R}$. The cut-off and exponent parameters
as well as the scale parameter $s_0$ of the spectrum of spectator states
are chosen to fit the charged-lepton and neutrino data for the
structure functions $F_2$ and $xF_3$ at average momentum transfer
$Q^2$ between 5 and 10\,GeV$^2$ (for more details see ref.\cite{KPW94}).

A comment on the scaling violation is in order.
In our approach we fix the parameters of the spectrum of spectator
states to describe the measured structure functions at some fixed
$Q^2$ and therefore effectively incorporate the $Q^2$ evolution
effect in the phenomenological parameters of the spectrum.
A more consistent approach which
accounts for the scaling violation effect would be to choose a low
resolution scale, $Q^2\sim 1$\,GeV$^2$, and fix the parameters of the
spectrum at this scale. Then at larger $Q^2$
the parton distributions are obtained by applying the $Q^2$
evolution equations. This program in fact has been attempted in
ref.\cite{KMWW96}.

\section{Nuclear Shadowing}
\label{shadow}

We now turn to discussion of DIS on nuclear targets in the small $x$
region.
The physical origin of nuclear corrections in the small $x$ region is
the coherent interaction the propagating $q\bar q$ pair with bound
nucleons.
In the parton model regime, which is discussed in this paper, the
$q\bar q$ pair is highly asymmetric in the target rest frame, i.e.
one of the pair constituents carries practically all the transferred
momentum.  It is commonly assumed that the fast quark only weakly
interacts with the target and begins to radiate soft gluons and
hadronize well after the nucleus. The other participant, the wee
(anti)quark, has a finite momentum and interacts in the nucleus with
normal hadronic cross-section.

It is instructive to compare different time and length
scales involved in this process.
The onset point of coherent nuclear effects $x_o$ is usually estimated by
the comparison of the propagation length $L\sim (2Mx)^{-1}$ of the $q\bar
q$ pair with the averaged distance between bound nucleons in the
nucleus.  If we take for estimates the averaged nuclear density
$\rho=0.17\,$fm$^{-3}$ in the central nuclear region then
$x_o\approx0.1$.  We note however that this is only a kinematical
condition which allows for coherent interactions of the propagating
$q\bar q$ pair with several nucleons for $x<x_o$.
It is quite clear that the region of developed
nuclear shadowing
should depend on (anti)quark-nucleon interaction.
Indeed if $\sigma$ is the averaged (anti)quark-nucleon total cross
section then the (anti)quark mean free path in the nucleus
$l=(\rho\sigma)^{-1}$.  The multiple scattering effect becomes
significant when the coherence length $L$ exceeds the (anti)quark
mean free path in the nucleus $l$.
This happens when $x < \rho\sigma/2M\approx 0.02$,
where
we have taken for estimates $\sigma=\sigma_{pp}^{tot}/3\approx13\,$mb.
%
%
In this case the main part of the incoming flux of $q\bar q$ pairs is
absorbed at the nuclear surface in the layer with the thickness of
order $l$ and thereby the inner nucleons are screened.

In order to quantify the discussion we consider the quark and antiquark
nuclear scattering amplitudes in forward direction, $T_{qA}$ and $T_{\bar qA}$,
and apply the Glauber-Gribov
multiple scattering expansion \cite{Glauber,Gribov}
assuming that formalism can be taken over to off-shell interaction,
\begin{eqnarray}
\label{dT}
T_A &=& A\, T_N + \delta^{(2)} T_A + \ldots  \\
{i\,\delta^{(2)}T_A(s,k^2)\over 2s} &=& A(A-1)
\left({i\,T_N(s,k^2,t) \over 2s}\right)^2
\int\limits_{z'>z}\!\!\!d^2\bm{b}\,dz\,dz'\,
e^{i q_\|(z-z')} \rho^{(2)}(\bm{b},z,z'). \nonumber
\end{eqnarray}
Here we have written explicitly only the first and the second terms in
the expansion. The first term accounts for the incoherent scattering from bound
nucleons while the second one is a correction due to the double scattering.
The dots denote higher order scattering terms.
The notations are as follows: $k=(k_0,\bm{0}_\perp,k_z)$ is the momentum
of incoming off-shell (anti)quark,
$q_\|=(m_q^2-k^2)/2|k_z|$ is the longitudinal momentum transfer
in the scattering of the (anti)quark from
the nucleon, $t=-q_\|^2$, $s$ is the (anti)quark-nucleon
center-of-mass energy squared
and $\rho^{(2)}(\bm{b},z,z')$ is the
two-nucleon density matrix.
In our estimates we neglect effects due to inelastic excitations of the
projectile particle in intermediate states as well as the effects
of nucleon-nucleon correlations in the nucleus
and assume that $\rho^{(2)}$ factorizes into two
one-particle densities normalized to the unity,
$\rho^{(2)}(\bm{b},z,z')=\rho(\bm{b},z)\rho(\bm{b},z')$.  In the
numerical evaluations we use the Gaussian shape for the densities
$\rho(\bm{b},z)$ with the scale parameter fixed by the nuclear
root-mean-square radius.

We use \eq{dT} to calculate (anti)quark-nuclear scattering amplitudes
and then evaluate nuclear corrections to the quark $q(x)$ and antiquark
$\bar q(x)$ parton distributions in the isoscalar nucleon
by Eqs.(\ref{qII},\ref{qbarII}).
We keep the terms up to the 4th-order in the multiple scattering
series (\ref{dT}).
In Fig.3 we plot the ratios $R_2=F_2^A/A F_2^N$ and $R_3=F_3^A/A F_3^N$
for the calcium nucleus
for the structure functions averaged over neutrino and antineutrino
fluxes,
$F_2=(F_2^\nu+F_2^{\bar\nu})/2$ and  $F_3=(F_3^\nu+F_3^{\bar\nu})/2$
(the structure functions are evaluated for the isoscalar nucleon).
%
The double scattering correction is negative%
\footnote{
This is well seen from \eq{dT} if we neglect
the effect of $q_\|$ as well as a small
real part of the scattering amplitudes.}
and brings the main contribution to
the multiple scattering series.
However the share of the 3d-order and the 4th-order terms is also sizable,
about 25\% of the double scattering term in $F_3$ and about 15\% in the case
of $F_2$ for the calcium nucleus.
The sum of the 3d-order and the 4th-order terms is positive
and partially compensates the double scattering term.
Note also that the share of higher order terms increases in heavy
nuclei.

One can observe from Fig.3  that the nuclear shadowing effect for $F_3$
(i.e. for the valence quarks) is about 2 times stronger as the one for $F_2$
(i.e. for the sea quarks).
To clarify the underlying reason
for this factor of 2 we recall that the structure functions $F_2$ and
$F_3$ are determined by the sum and by the difference of the
antiquark and the quark scattering amplitudes respectively.  At high energy
the Pomeron pole dominates both the quark and
the antiquark scattering amplitudes.  The difference of the
amplitudes, which is determined by the $\omega$-Reggeon contribution,
is small as compared to their sum.  Since the double scattering
correction in \eq{dT} is bilinear in the quark amplitude, we observe
immediately that the relative correction to the difference of the
quark and the antiquark cross sections is two times as bigger as the
corresponding correction to their sum.%
\footnote{ The author is
grateful to N. N. Nikolaev for useful discussion on this point.}

%
\begin{center}
\begin{picture}(310,200)
\epsfxsize=100mm
\epsfbox{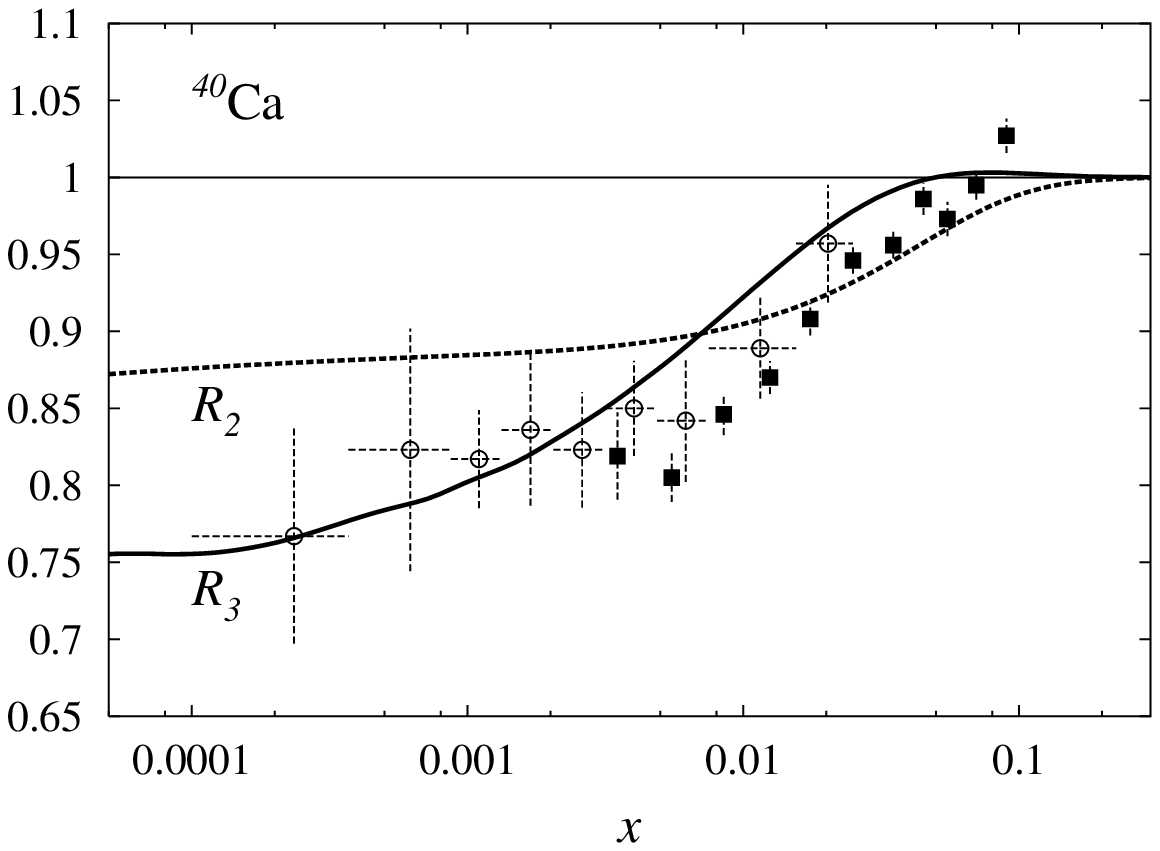}
\end{picture}
\\
\vspace{10pt}
{\small FIG.3}\\
\begin{minipage}{150mm}
{\vspace{10pt}\small
Ratios $R_2$ and $R_3$ 
calculated for the $^{40}$Ca nucleus 
for small $x$ and fixed $Q^2=5$\,GeV$^2$ (see text).
The data points for the ratio $R_2^\mu$
are from NMC \cite{nmc} (filled boxes)
and E665 \cite{e665} (open circles) experiments with muon beams.}
\end{minipage}
\end{center}

We found a sizable correction to the Gross-Llewellyn-Smith
sum rule for heavy nuclei due to nuclear shadowing.
For the calcium and iron nuclei the corrections are
$$
\int_{10^{-5}}^{0.3}dx\left(\frac1A F_3^A(x)- F_3^N(x)\right) 
\approx 
\begin{array}{l}
-0.11\ ({}^{40}\mbox{Ca})\\ -0.12\ ({}^{56}\mbox{Fe})
\end{array} .
$$
It is usually believed that the GLS sum rule should not be renormalized by
nuclear effects in the leading in $1/Q^2$ order,
because in this order the GLS integral counts the baryon number of the target.
It is challenging therefore to look for a mechanism which would
compensate a negative correction due to nuclear shadowing (`antishadowing').
Such an antishadowing may come from  $x>0.1$ region.
We note in this respect that
the Fermi-motion and nuclear binding corrections, which are of importance
at large $x$,
do not violate the GLS sum rule \cite{Kul98}.
A possible `compensating correction' could be due to off-shell modification
of the bound nucleon structure functions.

In Fig.3 for comparision
we put also the data points for $R_2^\mu$ for $^{40}$Ca nucleus
from NMC \cite{nmc}
and E665 \cite{e665} experiments with muon beams.
We note that data is taken for different $Q^2$
for every $x$-bin and, due to the nature of the fixed target experiments,
small $x$ is always correlated with small $Q^2$ (for events with
$x<10^{-2}$ presented in Fig.3, averaged $Q^2 < 1\,$GeV$^2$).
In the small $Q^2$ region where scaling is violated, the applicability of
the leading twist calculation is questionable.
We therefore need a model which provides a smooth transition from scaling to
nonscaling region.
In \cite{KPW94} we have considered a model which interpolates the structure
function $F_2^\mu$ (measured in muon induced reactions) between scaling and
nonscaling regime where the latter was described in terms of vector meson
dominance model \cite{PW90} (VMD).
It was found in the framework of this two ``phase model" that the VMD part
of $F_2$ is shadowed strongly than its scaling part.
This lead to the enhancement of the overall nuclear shadowing effect.
A similar observation has been done in a recent analysis of
nuclear shadowing
effect in neutrino induced DIS \cite{BLT98}.
In conclusion we notice the importance
of a small $Q^2$ analysis for the structure function $F_3$.
We recall in this respect that $F_3$ describes the
axial-vector--vector current transitions in the target under the
charged current and the studies of this effect may bring new insights
into the problem of chiral symmetry violation in nuclear environment.

\vspace{10pt}\noindent
The author is grateful to W. Melnitchouk and N. N. Nikolaev
for useful discussions and to the organizers of the Seminar for
warm hospitality.\\
This work is supported in part by the RFBR grant 96-02-18897.



\end{document}